\documentclass[%
 amsmath,amssymb,
preprint,%
]{revtex4-1}

\usepackage{graphicx}% Include figure files
\usepackage{dcolumn}% Align table columns on decimal point
\usepackage{bm}% bold math
\usepackage[utf8]{inputenc}
\usepackage[T1]{fontenc}
\usepackage{mathptmx}
\usepackage{color}

\begin{document}

%\preprint{AIP/123-QED}

\title[Acoustically driven Dirac electrons in monolayer graphene]{Acoustically driven Dirac electrons in monolayer graphene}

\author{Pai Zhao}
\affiliation{Center for Hybrid Nanostructures (CHyN), Department of Physics, Universit\"at Hamburg, 22761 Hamburg, Germany}

\author{Lars Tiemann}
\affiliation{Center for Hybrid Nanostructures (CHyN), Department of Physics, Universit\"at Hamburg, 22761 Hamburg, Germany}
%\email{lars.tiemann@physik.uni-hamburg.de.}

\author{Hoc Khiem Trieu}
\affiliation{Institute of Microsystems Technology, Hamburg University of Technology, 21703 Hamburg, Germany}

\author{Robert H. Blick}
%\homepage{https://nanoscience-hamburg.de/}
\affiliation{Center for Hybrid Nanostructures (CHyN), Department of Physics, Universit\"at Hamburg, 22761 Hamburg, Germany}

\date{\today}

\begin{abstract}
We demonstrate the interaction between surface acoustic waves and Dirac electrons in monolayer graphene at low temperatures and high magnetic fields. A metallic interdigitated transducer launches surface waves that propagate through a conventional piezoelectric GaAs substrate and couple to large-scale monolayer CVD graphene films resting on its surface. Based on the induced acousto-electric current, we characterize the frequency domains of the transducer from its first to the third harmonic. We find an oscillatory attenuation of the SAW velocity depending on the conductivity of the graphene layer. The acousto-electric current reveals additional fine structure that is absent in pure magneto-transport. In addition we find a shift between the acousto-electric longitudinal voltage and the velocity change of the SAW. We attribute this shift to the periodic strain field from the propagating SAW that slightly modifies the Dirac cone.
\end{abstract}

\maketitle

Probing two-dimensional electron gases (2DEGs) with acoustic sound waves is a well-established technique in condensed matter physics.~\cite{Steven1996} This technique, which is based on launching surface acoustic waves (SAWs) in the MHz- to GHz-range, is now heavily applied in optomechanics and quantum computing applications.~\cite{Jens2007,Martin2014,David2017,Per2019} In a way one can compare the interaction of electrons with a SAW to how surfers ride on a water wave. Hence, as an observer one expects a significant variation if the energy dispersion is switched from a conventional parabolic dispersion to the quasi-relativistic linear dispersion of graphene. In other words the surfers would be independent from the energy of the driving wave, while conventionally the velocity of sound along the surface of a medium is given by $v=\frac{\omega}{q}$, with $q$ being the wave vector and $\omega$ the frequency of the wave.  

In piezo-electric materials, the propagating sound wave is dynamically accompanied by an electrostatic potential wave that contributes to the transport of charge carriers as an acousto-electric current $I_{aec}$ \cite{Shilton1995}. The SAW amplitude attenuates exponentially into the bulk of the medium within one wavelength $\lambda$~\cite{Dustin2016, Lev2017}, which usually exceeds the depth of conventional 2DEGs. Experiments using SAW to study GaAs/AlGaAs heterostructures have been able to detect a wave velocity shift $\Delta v$, and attenuation, $\Gamma$, which both are functions of conductivity and reflect the conductivity oscillations in the quantum Hall regime or interactions and phase transitions\cite{Wixforth1986,Wixforth1989,Willett1990}.

In the current work we intend to probe if SAW-driven Dirac electrons show a different kind of response in the electro-acoustic current, as it is presumed by Thalmeier~\cite{Peter2010} and others~\cite{Minguez2018}. The coupling of the SAW piezoelectric field to the electron gas can attenuate the wave propagation whereas periodic strain fields can modify the Dirac cone in graphene. Thalmeier \textsl{et al.} predict that the wave-vector dependence of the longitudinal conductivity would reveal a Dirac to Schrödinger crossover. Graphene, however, is not intrinsically piezoelectric. Hence, SAWs must be launched in a piezoelectric medium in close contact to the graphene layer in order to study acousto-electric effects.  

%A surface acoustic wave is an elliptically polarized wave that propagates with the velocity $v=\frac{\omega}{q}$ 
%($q$: wave vector, $\omega$: frequency) of sound along the surface of a medium\cite{Steven1996}. Miniature %semiconductor devices using SAWs are found in both industrial applications and fundamental research, ranging from %daily telecommunications, life science and quantum information processing\cite{Per2019},\cite{David2017},
%\cite{Martin2014},\cite{Jens2007}. 

\begin{figure*}[ht]
\includegraphics[scale=0.35]{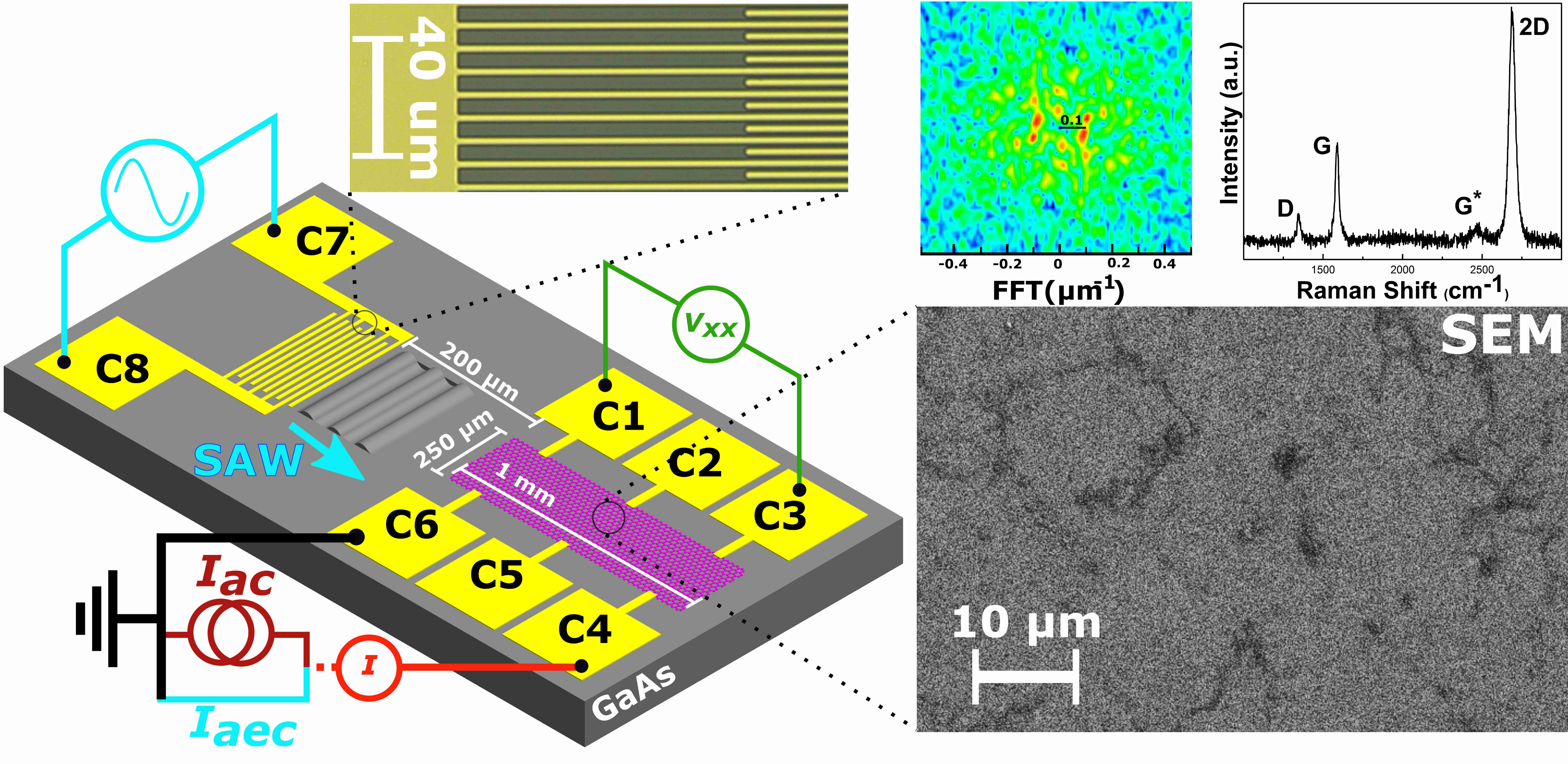}
\caption{\label{fig1}Sample design and sketch of the measurement setup: an IDT (top inset) with contacts C7 and C8 was fabricated on a GaAs semi-insulating piezoelectric substrate (grey). By applying an RF signal, a SAW (light blue) is launched that propagates through the graphene-Hallbar of 1 mm length (magenta honeycomb lattice). The right-hand lower panel shows a SEM image that highlights defects and boundaries existing in our CVD-graphene. The upper panel in the middle is the FFT analysis of these ripples and wiggles (blue represents minima, red maxima). The ratio of the G and 2D peaks in the Raman spectrum shown on the right-hand upper panel demonstrates the existence of a single layer of graphene.  A magnetic field is applied perpendicular to the sample (not shown).}
\end{figure*}

\begin{figure}[!hb]
	\includegraphics[scale=0.35]{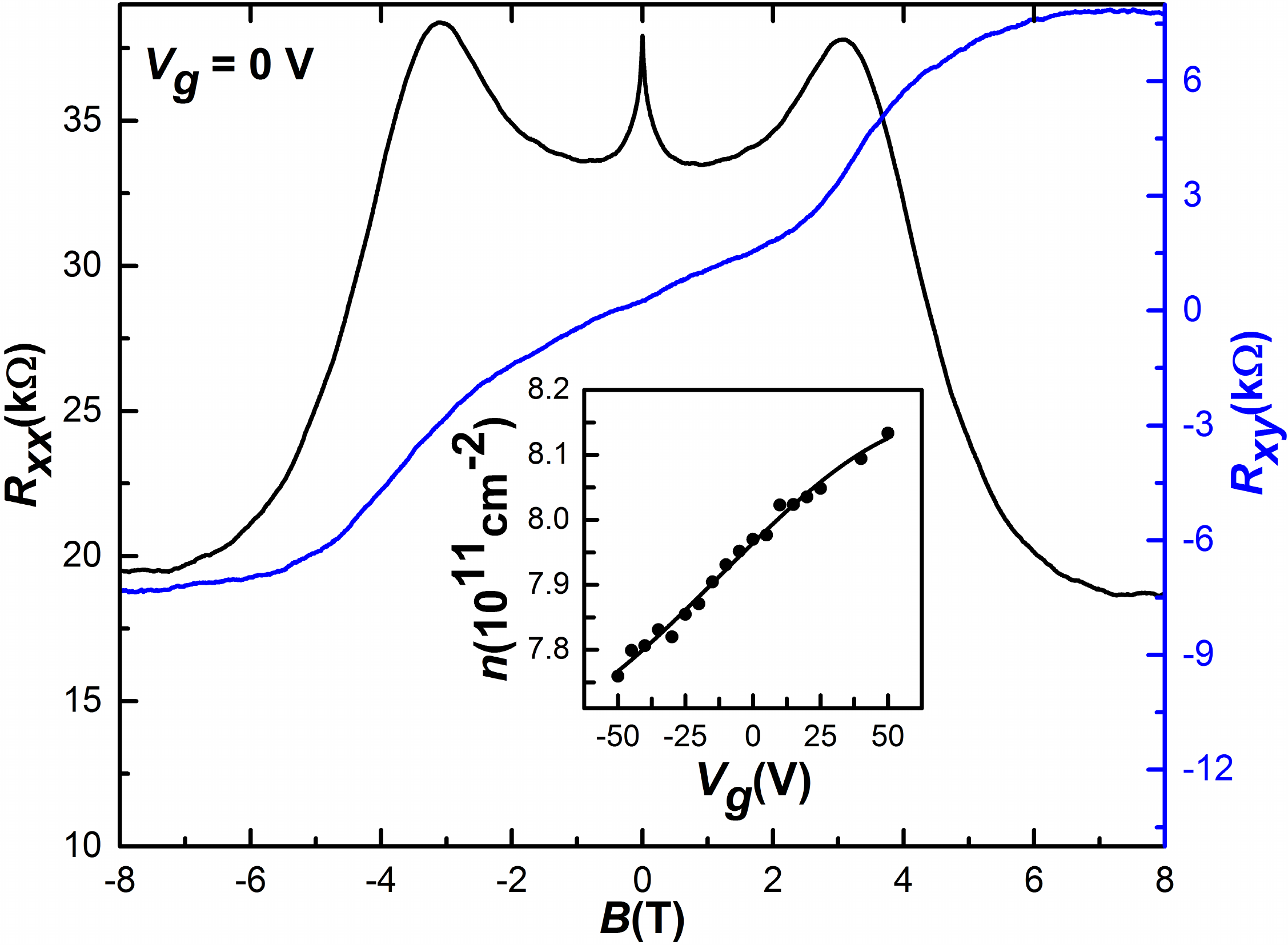}
	\caption{\label{fig2} Longitudinal resistance (black solid line) and Hall resistance (blue solid line) measured by passing a constant ac current through the graphene Hall bar at 4.2 K and $V_g$ = 0 V. Landau quantization is signaled by a Hall plateau for $|B|$ > 6 T. Inset: carrier density as function of back gate voltage.}
\end{figure}

In detail, we present a study of SAW in graphene in the $qd \ll$ 1 regime, where $d$ is the thickness of monolayer graphene, using a GaAs substrate as the piezoelectric medium. We calculate changes in the SAW velocity based on magneto-transport measurements obtained in a standard setup using a constant ac current. Our sample processing begins with electron beam lithography to define IDTs on top of a piezoelectric insulating GaAs substrate that launch SAWs. In a wet-transfer process\cite{Timothy2017}, monolayer CVD graphene is placed on the substrate. At a distance of 200 $\mu$m and in the propagation path of the SAWs, a Hall bar of 1 mm $\times$ 0.25 mm is fabricated by optical lithography. Deposition of Au/Ti is used for both the IDTs and ohmic contacts to the Hall bar. Prior to our measurements, the sample was thermally annealed in vacuum at 100 $^\circ$C for 48 h to reduce the amount of contaminants on the graphene surface. The sample configuration and the overall setup are schematically shown in Fig.~\ref{fig1}. The right-hand panels show a scanning electron microscopic image and a Raman spectrum of the graphene layer, indicating that we are indeed work on monolayer graphene.

Transport measurements are performed with SR830 lock-in amplifiers to detect currents and voltages. They also provide a reference to the signal generator for amplitude modulation (AM) of the radio frequency (RF) signal applied to the IDT\cite{Bruno2005, Heil1984}. In the current setup we work with one IDT firing at the graphene sample. All measurements are performed at 4.2 K. The graphene is characterized by standard magneto-transport using an ac current of $I_{ac}\approx$ 2 nA (red branch connected to C4 in Fig.~\ref{fig1}). Fig.~\ref{fig2} shows the resulting longitudinal resistance, $R_{xx}$, and Hall-resistance, $R_{xy}$, as function of the magnetic field, $\vec{B}$, at a back-gate voltage of $V_g$ = 0 V. The formation of a plateau in $R_{xy}$ for $|B|$ > 6 T signals Landau quantization. From the Hall resistance, we can deduce the intrinsic electron density of $n$ = 7.9$\times$ $10^{11}$ cm$^{-2}$ and the mobility of $\mu$ = 4.15 $\times$ $10^{2}$ cm$^{2}$V$^{-1}$s$^{-1}$. The chip carrier back contact acts as an electrode for tuning the carrier density. The small capacitance of the substrate, however, limits the tunability of the carrier concentration to approximately 3$\%$ (inset Fig.~\ref{fig2}). Thus, all subsequent measurements are performed at $V_g$ = 0 V.

\begin{figure}[ht]
	\includegraphics[scale=0.35]{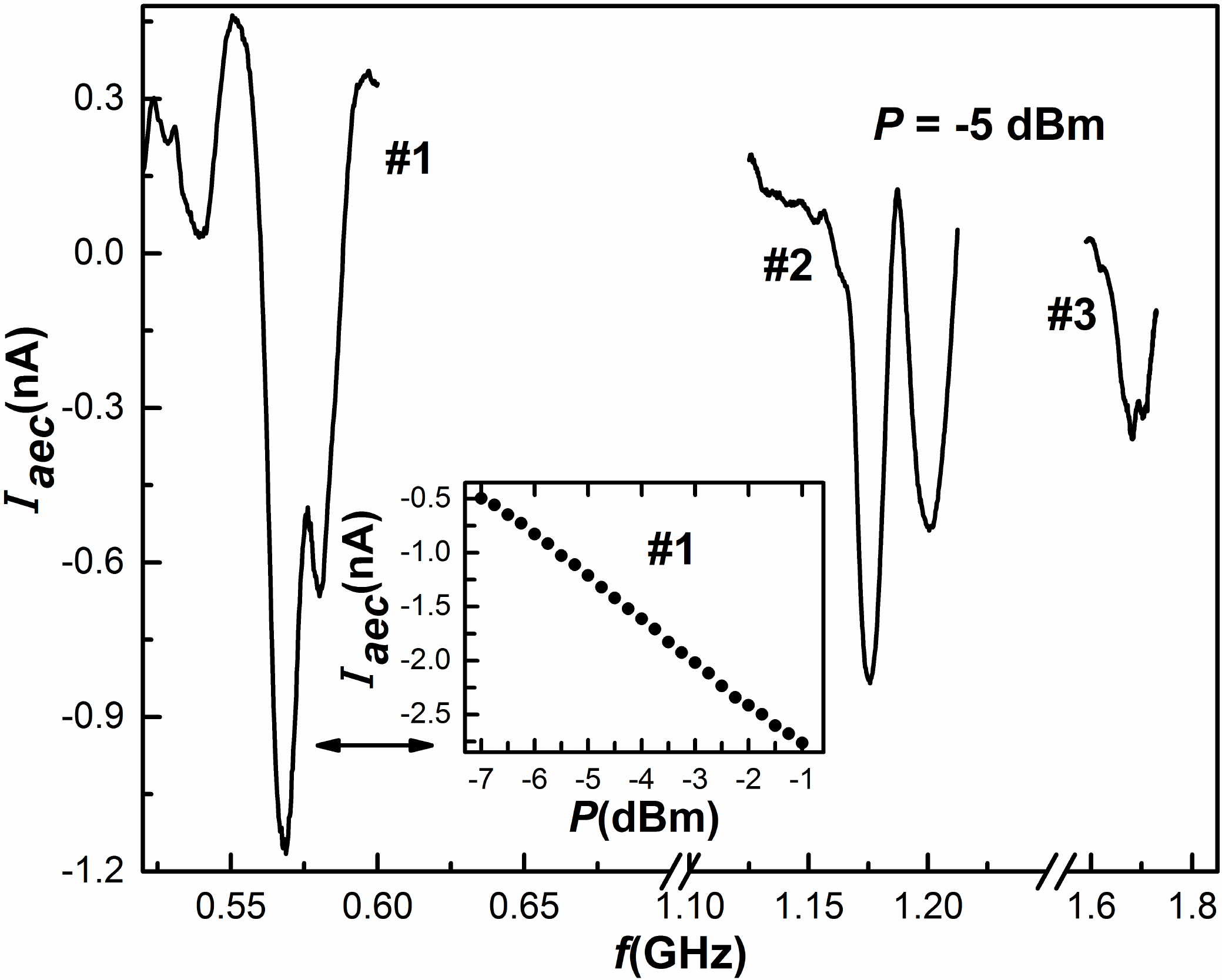}
	\caption{\label{fig3} Acousto-electric current as a function of frequency at -5 dBm with first harmonic resonance (\#1) at 571.5 MHz, second harmonic resonance (\#2) at 1.175 GHz and third harmonic resonance (\#3) at 1.684 GHz. Inset: linear power-dependence of the acousto-electric current for the first harmonic resonance of the IDT.}
\end{figure}

For tracing the SAW response, we disconnect the ac source and ground contact C4 (light blue branch in Fig.~\ref{fig1}). This is necessary in order to close the electrical circuit and enable a steady acousto-electric current. The SAW is launched by applying an AM RF power to the IDT. From the measurement of the acousto-electric current as a function of frequency (shown in Fig.~\ref{fig3}), we identify the (fundamental) first harmonic at 571.5 MHz, the second harmonic at 1.175 GHz, and the third harmonic at 1.684 GHz. For the first harmonic, $I_{aec}$ exhibits a linear dependence on applied RF power. Acousto-electric magneto-transport can therefore be performed with the same current amplitude of $I_{ac} \approx I_{aec} \approx$ 2 nA at 0 T as in the previous magneto-transport experiments using an ac signal.

The oscillating electrostatic potential that accompanies the propagation of a SAW in a piezoelectric medium depends on the screening capabilities of mobile carriers. The SAW velocity is thus a function of the conductivity and can be expressed as

\begin{equation}\label{eq1}
\frac{\Delta v}{v} = \frac{K^2_{\textsl{\normalfont eff}}}{2}\cdot\frac{1}{1+(\sigma_{xx}/\sigma_M)^2}
\end{equation}

with

\begin{equation}\label{eq2}
\sigma_M = v\epsilon_0(1+\epsilon_G)
\end{equation}

\begin{equation}\label{eq2}
\sigma_{xx} = \frac{\rho_{xx}}{\rho_{xx}^2+\rho_{xy}^2}
\end{equation}

$K_{\textsl{\normalfont eff}}^2$ = 0.06$\%$ is the piezoelectric coupling coefficient for GaAs, which we use as the piezoelectric substrate, and $\epsilon_G$ is the dielectric constant of graphene~\cite{Elias2011}. The longitudinal resistivity, $\rho_{xx}$, and Hall resistivity, $\rho_{xy}$, are obtained from magneto-transport in the standard ac signal as shown in Fig.~\ref{fig2}.

In our sample, interfacial van der Waals bonds work as a bridge between the monolayer graphene and the GaAs surface, which will also determine the coupling of the SAWs to the graphene. To our knowledge, no reports exist on the exact value of the van der Waals force between graphene and GaAs. However, for graphene on SiO$_2$ both density functional theory and experimental methods\cite{Zachary2010, Steven2011, Wei2014, Deji2017} yield adhesion energies at the interface of the order of $\sim$ 100 mJ/m$^2$ or van der Waals forces of the order of $\sim$ 100 MPa, respectively. Here, we assume similar orders of magnitude for the GaAs interface.

\begin{figure}[!ht]
\includegraphics[scale=0.35]{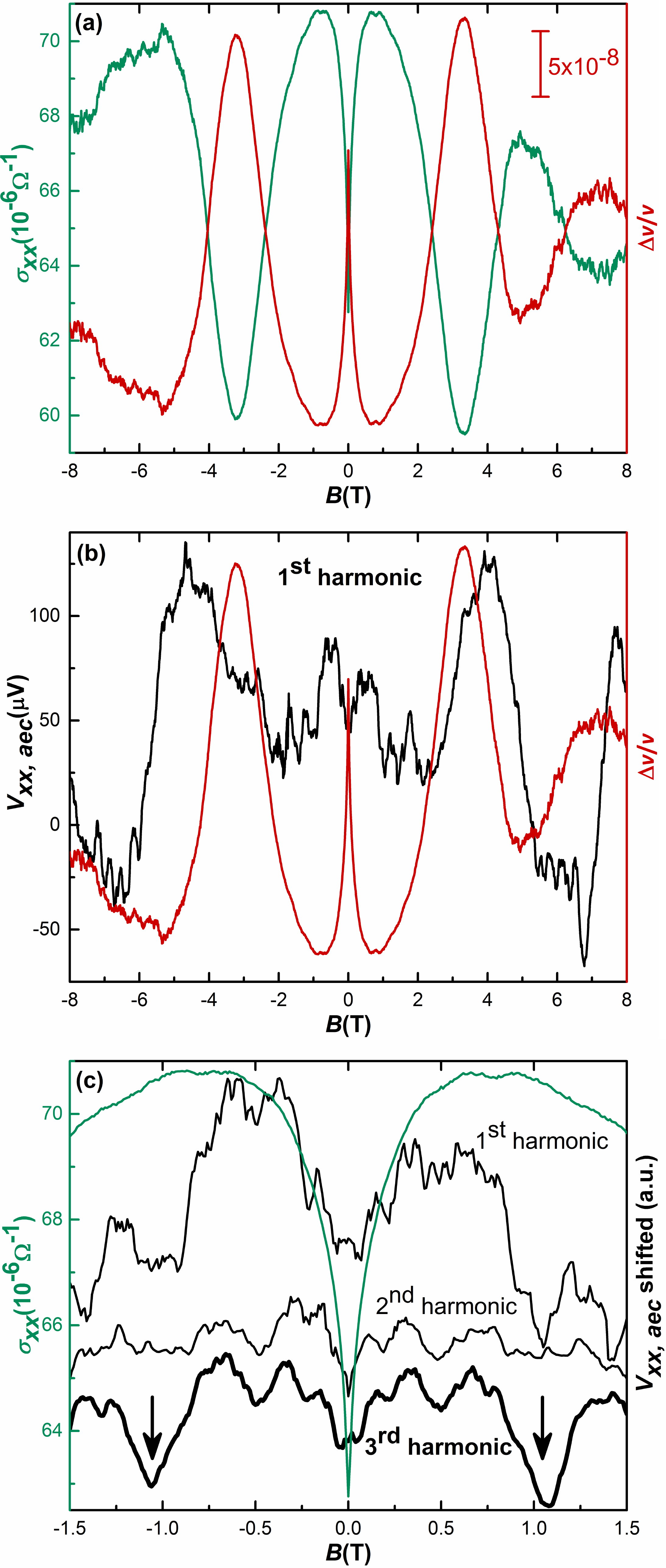}
\caption{\label{fig4}(a) SAW velocity change (right-hand axis) calculated from the conductivity (left-hand axis) measured by standard magneto-transport shown in Fig.~\ref{fig2}. (b) Longitudinal voltage oscillations (left-hand axis) generated by the acousto-electric current at the first harmonic and oscillations of SAW velocity (right-hand axis) as calculated in sub figure (a). (c) Longitudinal voltage induced by the acousto-electric current at the IDT’s first to third harmonic (see text for further details). The curves were shifted for clarity.}
\end{figure}

Graphene is known to exhibit rippling with a periodicity of a few nanometers when it conforms to an underlying substrate\cite{Fasolino2007}. This would give rise to local decoupling from the substrate. However, as the SAW wavelength exceeds this rippling by several orders of magnitude and in light of strong van der Waals forces, we assume Eq.~(\ref{eq1}) remains valid in our case.
Fig.~\ref{fig4}(a) shows the SAW velocity change calculated from the conductivity determined by standard magneto transport following Eq.~(\ref{eq1}). With decreasing conductivity, $\frac{\Delta v}{v}$ increases until a maximum is reached. This behavior was also observed in conventional 2DEGs in GaAs heterostructures and confirms the applicability to our graphene system and the values of the effective piezoelectric coupling, $K_{\textsl{\normalfont eff}}^2$, and characteristic conductivity, $\sigma_M$. Hence, it already appears likely that the charge carriers behave similarly to conventional 2DEGs in semiconductors.

Fig.~\ref{fig4}(b) shows the longitudinal voltage drop generated by the acousto-electric current. We find that the pronounced maxima in the longitudinal voltage are shifted with respect to maxima in the velocity change\cite{Steven1996}. We also observe an oscillatory fine structure in the acousto-electric longitudinal voltage drop over the entire magnetic field range, which does not appear in the conventional magneto-transport using a constant ac signal. Furthermore, we observe similar features at other harmonics. This fine structure also appears at the other IDT’s harmonics as shown in Fig.~\ref{fig4}(c).

While the fine structure is reminiscent of commensurability oscillations\cite{Weiss1989} that appear when the cyclotron orbits match periodic background potentials, the magnetic length, $l_B=\sqrt{\hbar/eB}$, at 10 mT is already much smaller than the SAW wavelength. Thus, it is impossible to reconcile the field dependence of the cyclotron orbits with our data. We assume that the fine structure is an interference effect related to the disordered nature of our sample. As shown in the inset of Fig.~\ref{fig1}, the CVD graphene appears to have ripples and wiggles with a periodicity of 1 $\mu$m and 10 $\mu$m. The large number of grains and folds give rise to a strongly non-uniform density distribution. A SAW propagating through the area covered by the graphene will thus experience a non-uniform attenuation and may branch off into a multitude of secondary waves; this is comparable to universal conductance fluctuations, which originate from the interference between the trajectories of all electronic paths in strongly disordered materials~\cite{Umbach1984,Lee1985}. The fine structure we observe may thus reflect the interference of all these secondary SAW paths.

The shift between the calculated velocity change and the measured acousto-electric voltage as shown in Fig \ref{fig4}(b), on the other hand, seems to be indicative of corrugation strain induced by the dynamically propagating SAW. The shear force generated by the SAW propagation is of the order of $\sim$ 0.1 MPa for graphene on polymer substrates\cite{Tao2014}. The shear strength of graphene on GaAs is potentially bigger since the van der Waals force of graphene on GaAs is larger than on polymer\cite{Deji2017}. Shear strain in graphene is predicted to induce strong gauge fields and would also affect the Landau quantization\cite{Sasaki2005, Guinea2010}. Therefore, strain is a strong candidate to explain the shift between the calculated velocity change and the measured acousto-electric voltage as shown in Fig. ~\ref{fig4}(b).

Closer inspection of Fig.~\ref{fig4}(c) seems to resolve periodic features in the velocity change $\frac{\Delta v}{v}$, while the conductance is mostly featureless. Among the different harmonics, one resonance seems to be very pronounced, i.e., for the third harmonic a peak structure at $\pm$ 1 T is resolved, which resembles weak localization at $B$ = 0 T. Combined with the wiggles and ripples of the graphene layer, we assume that the SAWs probe this disordered electronic system.
 
In summary, we presented a technology to study the carrier dynamics of Dirac electrons in graphene at high magnetic fields and low temperature using acousto-electric currents. We demonstrated the coupling between the propagating SAW and the electronic system through the attenuation of the SAW depending on the conductivity of the graphene. We can state that the electrons in this Dirac material behave mostly classical. However, it appears that there exists a fine structure which cannot be explained with a classical electron dispersion. This method is easily transferable to other van der Waals materials and other piezoelectric substrate even at room temperature.\newline
 
We acknowledge support by the Partnership for Innovation, Education, and Research (PIER). We also thank the Excellence Cluster Center for Ultrafast Imaging (CUI) of the Deutsche Forschungsgemeinschaft (DFG) for support under contract number EXC-1074. In particular, we would like to thank Steve H. Simon and Lev G. Mourokh for fruitful discussions.

\nocite{*}
\bibliography{aipsamp}

\end{document}